# Anomalous Crystallinity and Magnetism in Chemically Disordered Coherent Heterostructures

Saeed S. I. Almishal[1], Sai Venkata Gayathri Ayyagari[1], Aaron Pearre[2], Pat Kezer[3], Matthew Furst[1], Christina M. Rost[4], Binghai Yan[2], Nasim Alem[1], Timothy Charlton[5], Zhiqiang Mao[2], John T. Heron[3], and Jon-Paul Maria[1]

[1]*Department of Materials Science and Engineering, The Pennsylvania State University, University Park, PA 16802, USA*
[2]*Department of Physics, The Pennsylvania State University, University Park, PA 16802, USA*
[3]*Department of Materials Science and Engineering, University of Michigan, Ann Arbor, MI, 48109 USA*
[4]*Department of Materials Science and Engineering, Virginia Polytechnic Institute and State University, Blacksburg, VA 24061, USA*
[5] *Neutron Science Division, Oak Ridge National Laboratory, Oak Ridge, TN 37831, USA*

Corresponding author: Saeed S. I. Almishal ssa5409@psu.edu

## Abstract

High-entropy oxide (HEO) thin films uniquely superimpose exceptional chemical disorder with exceptional crystalline quality and coherence – an intersection we term *anomalous crystallinity* that arises from coupled structural, chemical, and valence degrees of freedom unique to the entropy-stabilized condition. Here, we demonstrate unexpected and predictive control of this state using formulation, epitaxial constraints, and kinetic arrest of metastable macrostates. Specifically, aliovalent cation substitutions, tightly controlled substrate temperatures, and conditions favoring significant adatom kinetic energy, can program the out-of-plane lattice parameter of coherent rock salt HEOs while preserving in-plane epitaxial pinning to MgO. Lattice strains exceeding 5% can be stabilized in multilayer heterostructures using this approach, where $3^+$ cations compensated by cation vacancies predominate the defect chemistry landscape. We highlight the exemplar (Sc,Mg,Co,Ni,Cu,Zn)O/(Cr,Mg,Co,Ni,Cu,Zn)O (JSc/JCr) system where Sc and Cr substitution into the rock salt structure produces pseudomorphic heterostructures between individual antiferromagnets, sustaining an exceptional 5.5% unit-cell volume and enabling abrupt interfaces across which the Co valence switches from mostly $2^+$ to an even $2^+/3^+$ mixture. These unprecedented valence interfaces are accompanied by a 2× exchange bias boost compared to single-layer constituents, that could be attributed to enhanced uncompensated spins in the layers themselves or around the buried JSc/JCr interface. These results establish pseudomorphic valence interfaces with anomalous crystallinity as a source of new magnetic macrostates that host emergent magnetic and spintronic functionality.

## Introduction

Synthesis kinetics grant access to many metastable macrostates, and thus atomic and electronic configurations, in high-entropy oxides (HEOs) by regulating the entropy-enthalpy exchange upon cooling from high temperatures or high kinetic energies[1–6]. By quenching eV-scale adatom kinetic energies ($10^4$–$10^5$ K effective temperatures) that are common to many physical vapor deposition methods on a comparatively cold substrate, nanosecond thermalization times can kinetically arrest far-from-equilibrium solubilities, coordination environments, and valence states that equilibrium synthesis routes cannot attain[7–10]. This rapid quenching yields films that are highly chemically disordered but with exceptional crystalline fidelity and unprecedented solvation power[10–12]. We refer to this unique combination as *anomalous crystallinity* – a hallmark of entropy-assisted and kinetically arrested epitaxial and metastable HEO thin films (Supporting Information Note 1).

Pulsed laser deposition (PLD) provides two critical avenues to the anomalously crystalline state. *First*, the high effective temperatures possible in laser plasmas (and unachievable for bulk synthesis) generates an enormous -$T$S term at the adatom arrival moment that favors the high entropy, high symmetry, and uniformly mixed state[7,8]. *Second,* the pulsed nature of PLD allows unique mediation of the cooling process as the rest time determines the surface-diffusion interval before the next layer arrives to freeze the subsurface material, or at least dramatically delay further relaxation. Additionally, laser fluence, deposition pressure and working distance regulate adatom arrival energy while substrate temperature governs surface and subsurface diffusion rates[6,7,10,11]. This combination can access a vast structural, microstructural, and defect chemical phase space in HEOs that may include unstable and metastable materials[6,10]. Substrate temperature has been shown to be particularly influential in this process, particularly in pseudomorphic rock salt HEOs[2,11]. In MgCoNiCuZnO grown on MgO (henceforth J14) for example, increasing the substrate temperature from 350°C to 375 °C expands the out-of-plane lattice parameter by ≈3.15 % while retaining exquisite crystalline quality and texture, and without generating misfit dislocations[2,11]. A Co valence shift from $Co^{3+}$ to $Co^{2+}$ produces this expansion due to the significantly larger $Co^{2+}$ ion radius[2,11]. While both structures are metastable, the lower temperature material is farther from equilibrium, and the modest substrate temperature boost allows relaxation into that lower enthalpy macrostate that host $Co^{3+}$.[2,6,11]

Our experiments show that the $Co^{3+}$ to $Co^{2+}$ valence exchange and the abrupt unit cell volume expansion in J14 thin films is driven by a thermodynamic transition between metastable states, this perspective implies that multiple experimentally accessible thermodynamic boundary conditions can influence and likely tune it. Furthermore, the structural degrees of freedom available to a high entropy parent phase like J14 will likely impart a significant and potentially gradual tunability through temperature, oxygen pressure, epitaxial strain, and formulation combinations during synthesis[2,6,11]. We know that strong structural and valence changes are possible through these transitions, and we posit a parallel opportunity to exploit them for property engineering[2,10,13–18]. To do so we focus initially on magnetism because J14 sustains long-range antiferromagnetic order despite substantial chemical disorder and ~40% nonmagnetic cations[19,20]. This places it in a diluted but still-percolating antiferromagnetic regime known to stabilize domain states that significantly enhance exchange bias[21–23]. Fluid control of Co valence, compensating point defects, and molar volume may create paths to partial chemical and/or defect ordering to strengthen magnetic coupling and stabilize new magnetic states.

Three integrated investigations explore this hypothesis. First, we introduce two contrasting trivalent cations into J14: relatively large $Sc^{3+}$ and relatively small $Cr^{3+}$. These expand the lattice constant tunability to over 7% with temperature modulation while maintaining exceptional crystalline quality. Second, we exploit this tunability in coherent bilayer heterostructures with valence interfaces. Third, we demonstrate how these heterostructure interfaces produce unexpected and unconventional magnetic states and we use polarized neutron reflectometry (PNR), to depth profile the interface-enabled magnetic structure. Collectively, we establish entropy-mediated anomalous crystallinity as a growth-tunable platform for coherent heterostructures and magnetic exchange-bias-based devices.

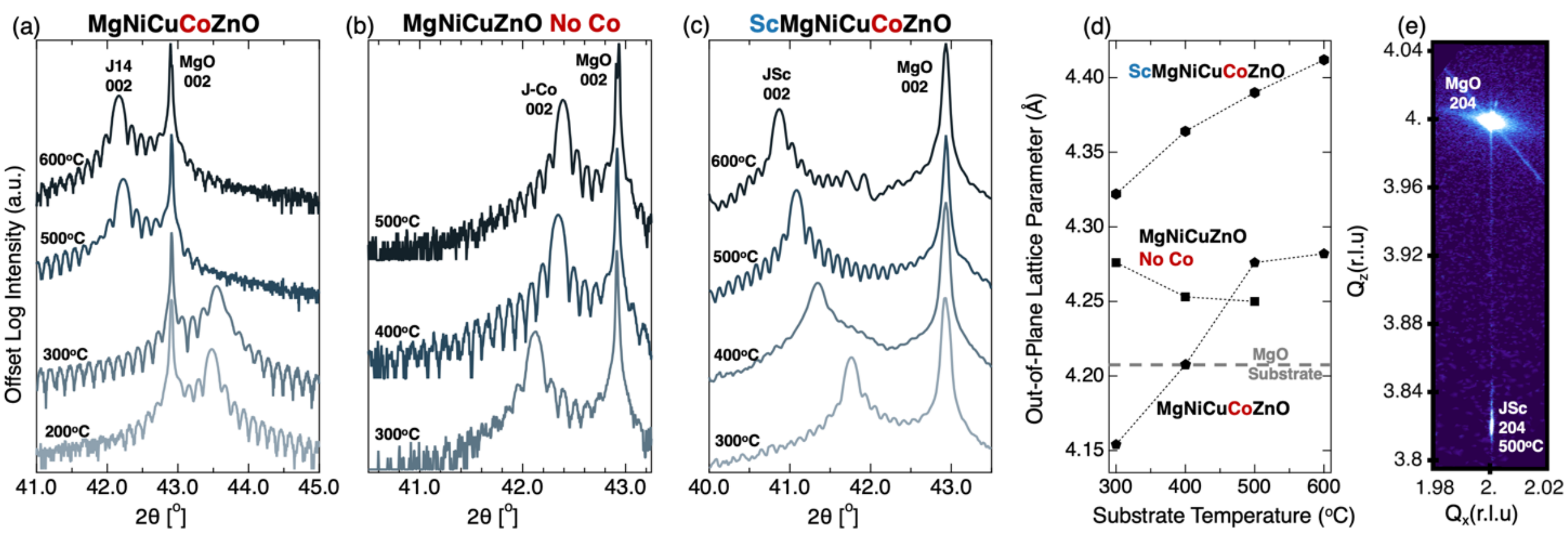

**Figure 1 | Temperature-dependent out-of-plane lattice parameter change in rock salt high-entropy oxide (HEO) thin films.** 2θ-θ X-ray diffraction scans around the (002) reflection for three rock salt oxide compositions grown on MgO(001) under 50 mTorr flowing $O_2$.: (a) the prototypical five-component HEO MgCoNiCuZnO (J14), (b) the Co-free quaternary MgNiCuZnO (J–Co), and (c) the Sc-containing six-component HEO ScMgCoNiCuZnO (JSc), deposited at different substrate temperatures as indicated. (d) Out-of-plane lattice parameter extracted from the film (002) reflection as a function of substrate temperature for J–Co, J14, and JSc films, illustrating composition-dependent out-of-plane lattice parameter changes under otherwise identical growth conditions. (e) Reciprocal space map of JSc grown at 500°C on MgO demonstrating this material family coherent pseudomorphic growth nature.

## Results and discussion

### Active unit cell volume and out-of-plane lattice parameter design

Figure 1 shows a collection of X-ray diffraction data highlighting the relative magnitudes and origins of anomalous strain possible in pure and modified J14 on MgO substrates. Figure 1(a) shows how increasing substrate temperature abruptly expands the *c*-axis while retaining coherence in all cases; this behavior has been attributed to Co[2,6,11]. X-ray patterns for J14 films with Co removed over the same temperature range are plotted in Figure 1(b) and show no lattice constant expansion supporting the Co valence redistribution hypothesis (Supporting Information Note 2).

We next add Sc as an equimolar sixth component to J14 – it is only ~2.8 % larger than the average J14 cation radius, closely size-matched to high-spin $Co^{2+}$, and only stable with $3^{+}$ valence – and prepare a JSc film series at substrate temperatures ranging from 300°C to 600°C. 2θ-θ XRD patterns in Figure 1(c) exhibit pronounced Pendellösung fringes, confirming uniform high-quality growth and smooth interfaces. In contrast to J14, JSc displays a smooth, monotonic increase in the out-of-plane lattice parameter from 4.32 Å to 4.41 Å (≈ 2.1 %) with increasing substrate temperature (Figure 1(d)). Stated alternatively, JSc at 300 °C already exhibits the high-temperature

J14 macrostate and higher temperatures further expand the lattice, with again, all samples pseudomorphic to MgO, this trend is highlight in the reciprocal space map (RSM) in Figure 1(e). We attribute these remarkable lattice constant changes to pseudo-strains to acknowledge likely contributions arising from substrate clamping, defect chemistry, and cation substitution, with defect chemistry and compensation playing a predominant role. If we estimate the Sc contribution using a rule of mixtures approach one estimates less than 0.5% contribution. However, $Sc^{3+}$ acts as a hard aliovalent dopant that promotes charge compensation through the defect reaction

$$\frac{1}{2}Sc_2O_3 \xrightarrow{J14} Sc_M^{3+^\bullet} + \frac{1}{2}V_M'' + \frac{3}{2}\ O_{O(g)}^{\times} \qquad \text{Eq. (1)}$$

This reaction will bias Co defect chemistry since both are linked by $[V_M'']$, the higher concentration forced by Sc will shift the Co defect reaction in Eq. (2) to the reactants side.

$$Co_{Co}^{2+^\times} \rightleftarrows Co_{Co}^{3+^\bullet} + \frac{1}{2}V_M'' \qquad \text{Eq. (2)}$$

More Co will be oxidized to the larger ionic radius $Co^{2+}$, locking the JSc films into a persistent pseudo-compressive strain state. Furthermore, higher temperatures will increase the $Co^{2+}$ fraction due to the increasing stability of lower valence states, which is experimentally observed.

We next examine the opposite perturbation by introducing Cr into J14. Due to its preferred $3^+$ valence, $Cr^{3+}$ is an aliovalent cation with a much smaller ionic radius (-15%) than the J14 average. $Cr^{3+}$ should also be compensated by metal vacancies as proposed for $Sc^{3+}$ thus favor a larger $Co^{2+}$ fraction and a similarly expanded out of plane lattice constant. However, their small radius will have the same contraction effect as $Co^{3+}$. Electron energy loss spectra (EELS) and XRD measurements reveal that Cr ions are overwhelmingly $3^+$, that 300°C J14 and 300°C JCr out of plane lattice constants are nearly identical, and that JCr grown at even higher temperatures retains the compressed unit cell volume and the exceptional crystalline quality and texture despite the large ion size mismatch. We note that at and above 500°C the JCr 002 peak shifts to higher 2θ values and new low intensity diffraction peaks with a doubled unit cell emerge. These may be associated with small volume fractions of spinel or a doubled rock salt oxide structure[6,9,24]. This behavior and the supporting EELS data are shown in Figure 2 (TEM details in Supporting Note 3). In fact EELS suggests an average Co oxidation state of ~$2.25^+$, this is higher than the predominantly $Co^{2+}$ state in J14 at the same temperature (500°C) and may result from the finite fraction of $Cr^{3+}$ that we suspect populates the minor second phase that emerges at 500 °C (More details on phase assignments are available in Supporting Information, Note 4). Ultimately, this J14 – JSc – JCr sample set demonstrates the unusual solubility of highly misfit cations (in both radius and valence), the unprecedented 7.8% lattice constant range that can be tuned with temperature and cation substitution, and the anomalous crystallinity, texture, and unexpected resilience against strain relaxation available to high entropy crystals. We now turn our attention to heterostructures that interface the extremes of this structural spectrum.

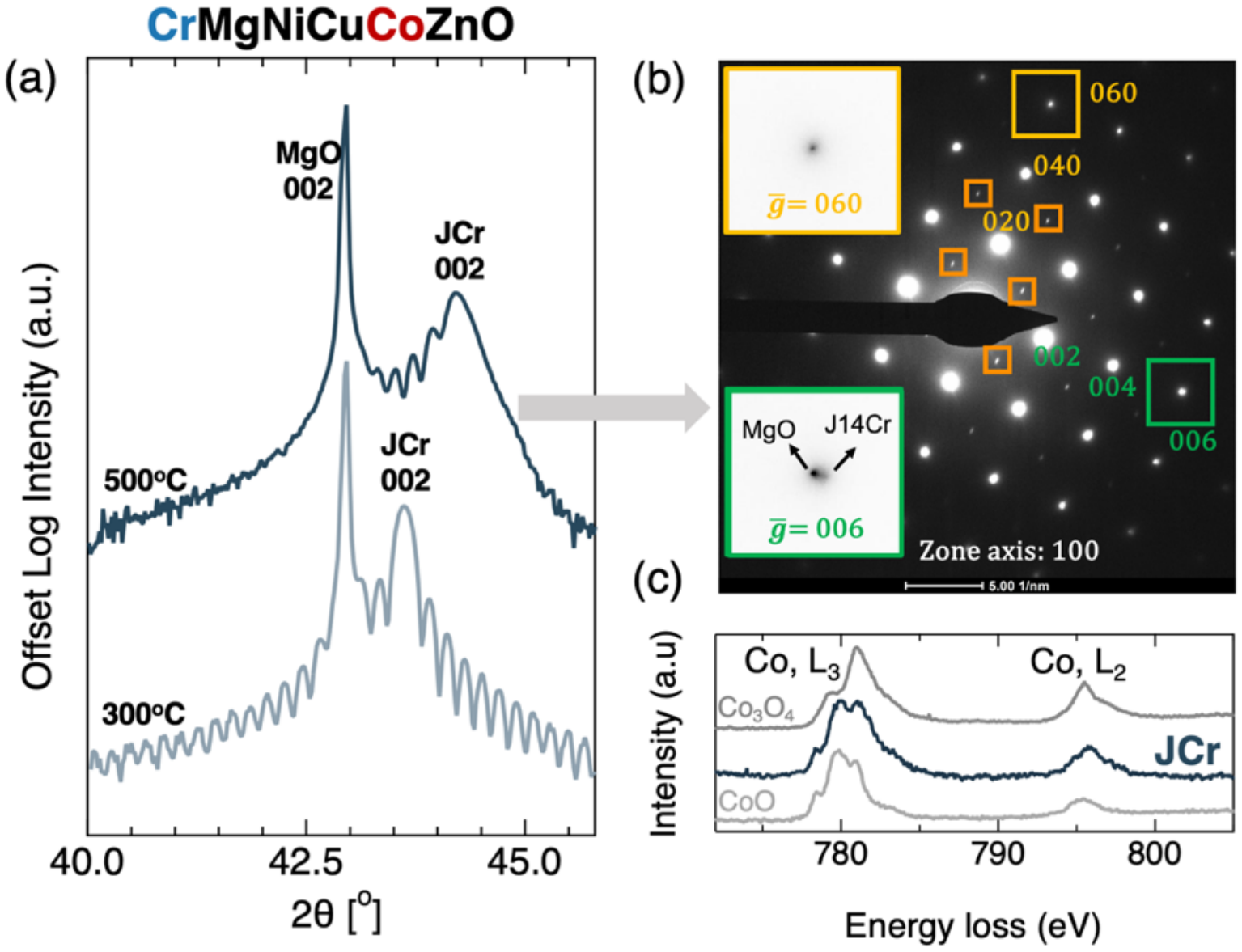

**Figure 2: Cr incorporation inverts the temperature trend by triggering a spinel-like ordering pathway.** (a) High-resolution 2θ–θ XRD scans of JCr (CrMgCoNiCuZnO) films grown at 300 °C and 500 °C. The 300 °C film exhibits exceptional crystalline quality, whereas the 500 °C film develops a pronounced right shoulder, consistent with the onset of spinel-like local ordering. (b) Selected-area electron diffraction (SAED) from the 500 °C film showing the coexistence of rock salt reflections and additional spinel-related diffraction spots (orange boxes). (c) Co $L_3/L_2$ EELS edges of the 500 °C JCr film compared with CoO ($Co^{2+}$) and $Co_3O_4$ (mixed $Co^{2+}/Co^{3+}$) reference spectra. The spectral shift toward the $Co_3O_4$ signature indicates $Co^{3+}$ enrichment and spinel-like ordering evolution at elevated temperature, corresponding to an average Co oxidation state of ~2.25+ from quantitative fitting.

## Engineering and kinetically arresting high-crystalline coherent heterostructures

We now harness the ~7.8% lattice constant (and point defect) tunability into two categories of coherent bilayered heterostructures: Type X and Type Z. Type X heterostructures feature two layers with pseudostrain that is the same sign with respect to the substrate while type Z heterostructures have one layer in pseudocompression (larger out-of-plane lattice parameter than the substrate) and one layer of pseudotension (smaller out-of-plane lattice parameter than the substrate).

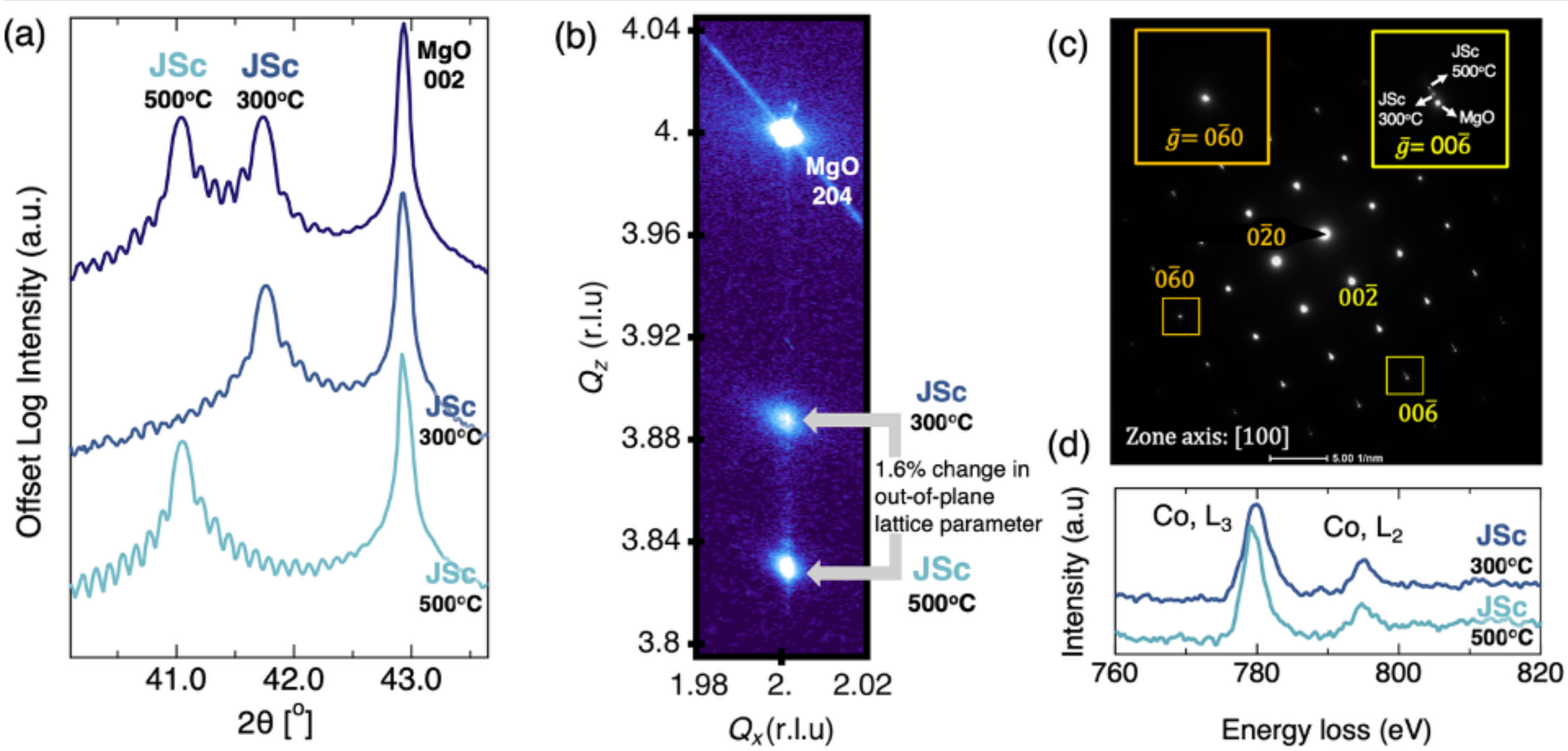

**Figure 3: Valence-gradient isochemical heterostructure (Type X) built from two coherently stacked JSc layers grown at 500 °C and 300 °C.** (a) 2θ-θ scans of the bilayer (top trace) compared to single-temperature JSc reference films (middle and bottom). The bilayer reproduces both c-axis spacings, confirming two distinct out-of-plane lattice parameters separated by ≈1.6 %. (b) Reciprocal-space map around the MgO (204) reflection. Two discrete film peaks share an in-plane lattice constant with the substrate yet differ along $Q_z$, demonstrating coherent growth. (c) Type X heterostructure plane-view SAED along the [100] zone axis, showing a single rock salt zone pattern. The single g= 060 reflection and three distinct intensity spots at g= 006, corresponding to the MgO substrate and the two film layers (JSc 300 °C and 500 °C), confirm coherent stacking without spinel-related diffraction features. (d) Layer-resolved Co-$L_{3,2}$ EELS. The 500 °C layer exhibits the lower-energy $Co^{2+}$-rich line shape, whereas the 300 °C layer shifts toward a mixed-valence profile, confirming a built-in valence gradient across an otherwise coherent interface. Spectra in **(a,d)** are vertically offset for clarity.

Our example type X heterostructure is a JSc bilayer grown at two substrate temperatures, the first epitaxial layer at 500 °C (out-of-plane lattice parameter +4.5 % vs. MgO) and the second layer at 300 °C (out-of-plane lattice parameter +2.0 % vs. MgO). Figure 3(a) shows high-resolution 2θ-θ scans with strong fringing and Figure 3(b) shows an RSM confirming pseudomorphic growth. The SAED in Figure 3(c) shows a single rock salt phase without secondary phases (Selected area region is in Supporting Information Note 5 Figure S6). Finally, layer-resolved EELS identifies a lower-energy, $Co^{2+}$-rich spectrum in the 500 °C layer and a mixed-valence signature in the 300 °C layer as shown in Figure 3(d), directly confirming the expected valence step across an otherwise coherent interface. Type X heterostructures of this type are isostructural and isochemical yet heterovalent: substrate temperature is the only difference separating them and it drives a sharp average oxidation state and defect equilibria transition.

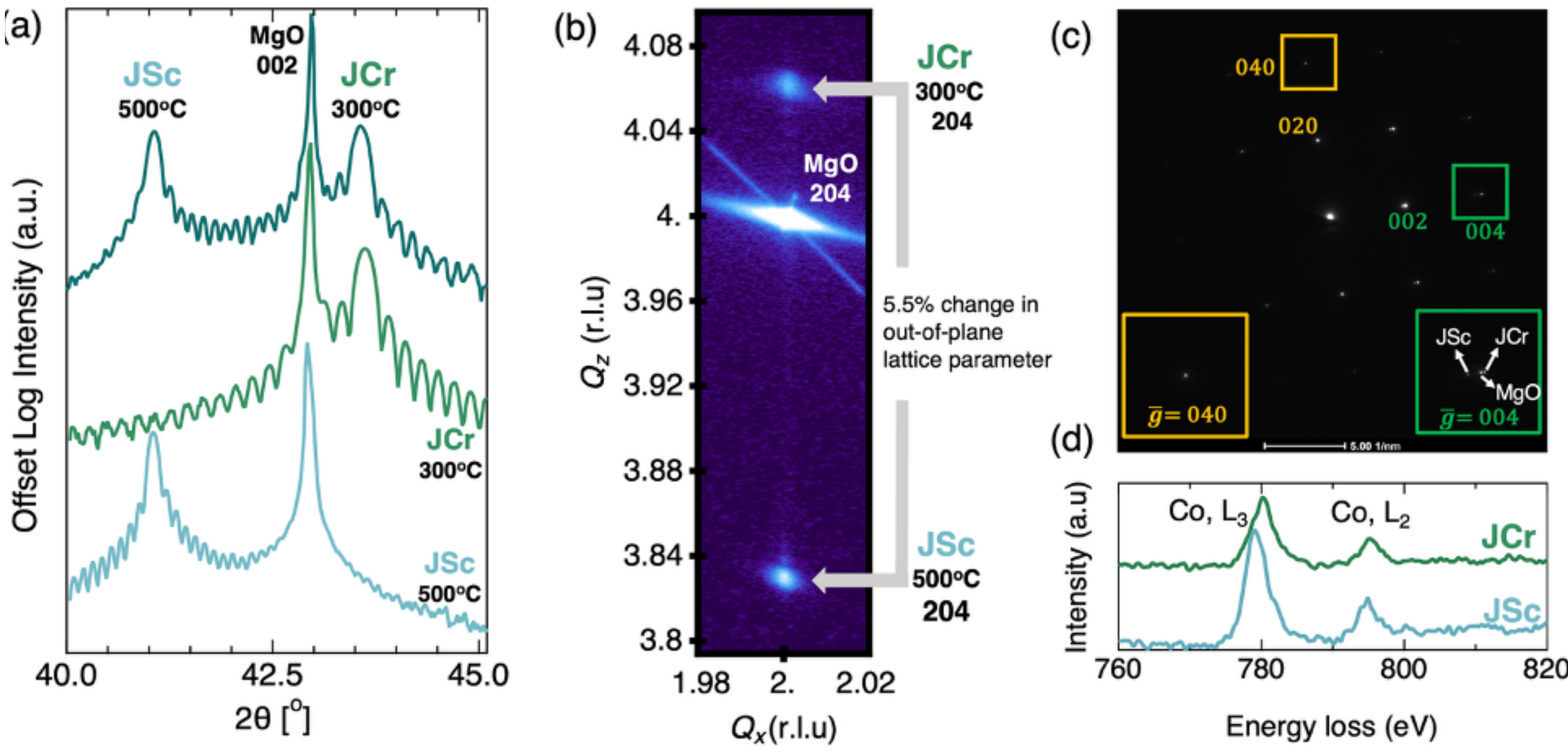

**Figure 4 | Large-contrast Type Z heterostructure built from a large JSc (500 °C) layer and a small JCr (300 °C) layer. (a)** 2θ-θ scans of the bilayer (top) alongside each of the JSc and JCr layers grown separately. Two distinct (002) peaks correspond to a ≈ 5.5 % c-axis difference. **(b)** Reciprocal-space map around the MgO (204) reflection. The two film spots lie on the same $Q_x$ value yet separate along $Q_z$, confirming coherent lateral matching and in-plane registry with MgO with different out-of-plane lattice contrast. **(c)** Plan-view SAED from the bilayer Type Z heterostructure. Only rock salt reflections are present (green boxes, *g* = 002/004); no spinel-specific spots appear, indicating that the low-T JCr layer remains spinel-free at deposition. **(d)** Layer-resolved Co-$L_{3,2}$ EELS. The JSc layer exhibits the lower-energy $Co^{2+}$-rich signature, whereas the JCr layer displays a slightly higher-energy, mixed-valence profile, establishing a built-in valence step. Spectra in **(a, d)** are offset vertically for clarity.

We next explore Type Z heterostructures in which the constituent layers experience opposing pseudo-chemical strain states This strain reversal can be maximized by combining 500 °C JSc layer with 300 °C JCr to engineer 5.5% pseudo-strain across a coherent and dislocation free interface (we note a second example is given in Supporting Information Note 6). Figure 4(a) shows high-resolution 2θ-θ scans with strong fringing similar to those of single layers and Figure 4(b) shows an RSM confirming pseudomorphic growth, both tensile and compressive layers adopt the substrate in-plane lattice parameter, with the SAED in Figure 4(c) showing only a single rock salt phase (Selected area region is in Supporting Information Note 5 Figure S6). Finally, layer-resolved EELS reveals that the JSc layer predominantly contains $Co^{2+}$, whereas the JCr layer exhibits an effective Co charge of approximately 2.5+. Together, these findings demonstrate structural and electronic transitions across interfaces only accessible in isochemical metastable high-entropy heterostructures with a sharp defect equilibria transition across remarkably strained interface (Strain maps of JCr/JSc stacked samples using geometric phase analysis (GPA) are available in Supporting Information Note 7).

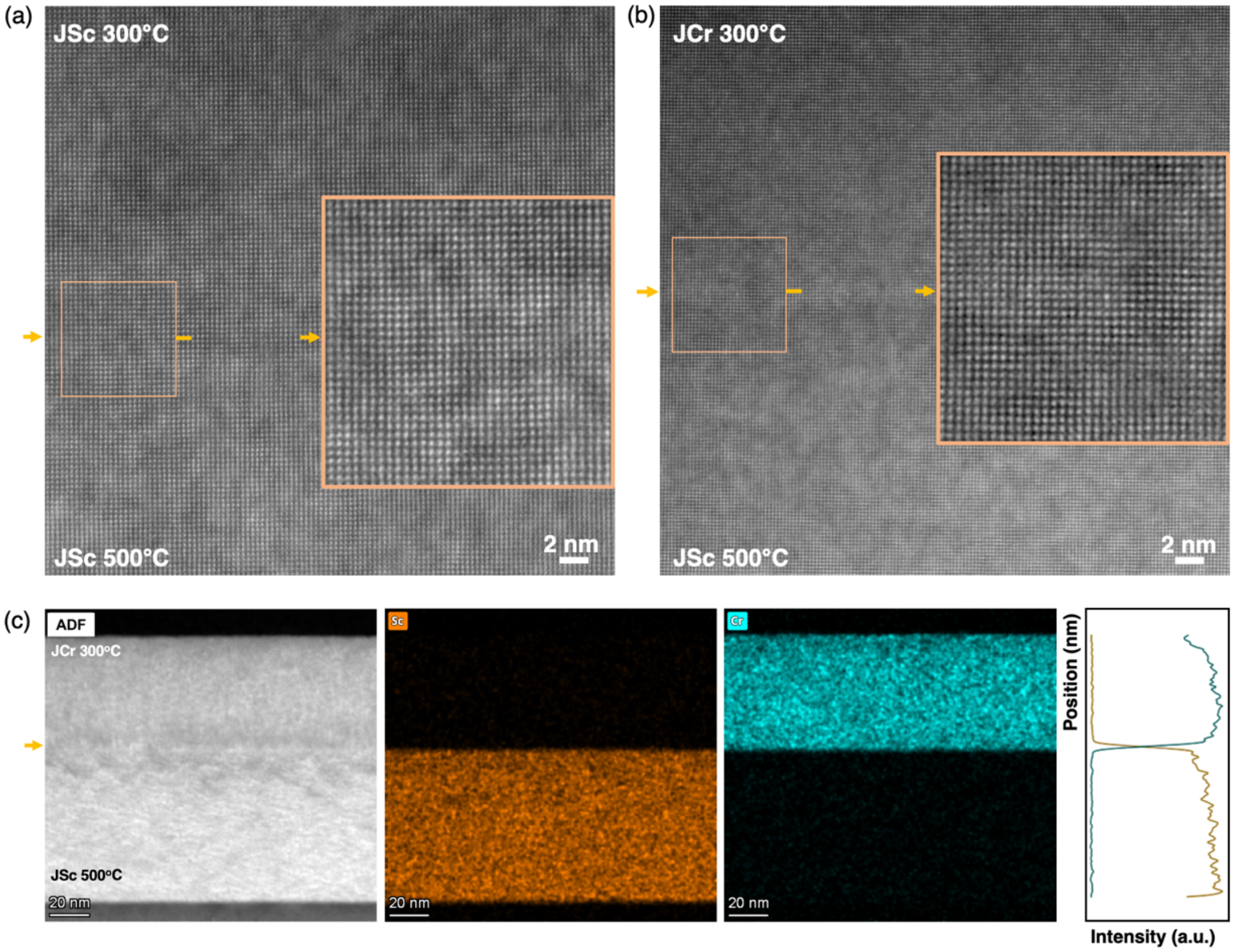

**Figure 5 | Atomic-scale high resolution imaging of Type X and Type Z heterostructures.** (a) High-resolution STEM image of a mild-offset (1.6%) Type X heterostructure (JSc/JSc), revealing atomically continuous, dislocation-free lattice coherence across the interface; the inset highlights the interfacial region. (b) High-resolution STEM image of the large-offset (5.5%) Type Z heterostructure (JSc/JCr), demonstrating atomically sharp chemistry and uninterrupted lattice continuity despite the substantial c-axis mismatch; the inset highlights the interfacial region, and (c) Annular dark-field (ADF) STEM image with corresponding EDS elemental maps across the JSc (500 °C)/JCr (300 °C) interface, showing an abrupt, chemically sharp boundary without detectable intermixing at the probed length scales. Line profiles confirm the smooth transition between Sc- and Cr-rich layers.

High-resolution STEM analysis of the heterostructures interfaces in Figure 5(a) and Figure 5(b) corroborate and confirm the remarkable structural continuity at the atomic scale, despite the significant associated interfacial strain of 1.6% and 5.5% for Type X and Type Z, respectively. Additionally, EDS maps for the Type Z heterostructure in Figure 5(c) confirm a clean, abrupt chemical interface between the Sc- and Cr-containing layers. With anomalous crystallinity and bilayer heterostructures established, we next investigate how these built-in gradients translate into functionality by considering magnetism.

## Magnetism across high-entropy chemically-disordered coherent interfaces

To probe the relationship between magnetic behavior and the structural and valence changes described above, we measured magnetic hysteresis in permalloy (Py)–capped films using a superconducting quantum interference device (SQUID, Quantum Design) magnetometer (See Methods). We focus on the Type Z heterostructure and its corresponding single-layer constituents because this architecture provides the largest contrasts in chemistry, valence state, and strain. We begin with the magnetization hysteresis data (M vs H) shown in Figure 6(a) for single-layer JSc and JCr thin films capped with 10 nm of Py, which allows direct visualization of exchange-bias behavior. After field cooling from room temperature to 10 K under an in-plane field of 2 T, both films exhibit clear horizontal loop shifts, indicating antiferromagnetic pinning at the Py/oxide interface. The JSc film grown at 500 °C shows an exchange-bias field of $H_{ex} = -42$ Oe, while the JCr film grown at 300 °C exhibits a larger bias of $H_{ex} = -54$ Oe. The enhanced exchange bias in JCr relative to JSc is consistent with a higher density of uncompensated interfacial spins. This can be attributed to partial conversion of magnetic $Co^{2+}(t_{2g}{}^5e_g{}^2)$ to magnetically inert $Co^{3+}$ $(t_{2g}{}^6e_g{}^0)$, combined with the presence of magnetically active $Cr^{3+}(t_{2g}{}^3e_g{}^0)$, which together introduce frustrated exchange pathways due to competing AFM and FM superexchange[11,17,25] and in turn enhance interfacial spin pinning at the Py/JCr FM/AFM interface.

The most striking magnetic response emerges in the JSc/JCr Type Z heterostructure also shown in Figure 6(a), which exhibits an of $H_{ex}=-113$ Oe – nearly twice that of either constituent layer alone under the in-plane field of 2T (Note that no magnetic hysteresis is observed in out-of-plane direction in Figure 6(a) and the inset shows that we reach in-plane saturation). This enhancement likely arises from cooperative coupling between the two pseudomorphic antiferromagnetic JSc and JCr layers, enabled by their exceptional chemical and valence interface. This enhanced Py pinning is unexpected given the distance separating the Py and the subsurface JSc layer, conventional understanding would predict identical behavior to Py/JCr. This invites a hypothesis where the "bulk" response of the JSc/JCr stack is modified by the JSc/JCr interface such that it can affect Py exchange bias, and while the specific origins are not mechanistically clear, valence , molar volume, and vacancy transitions across the coherent interface likely modify magnetic exchange and boost uncompensated spin densities leading to stronger permalloy pinning.

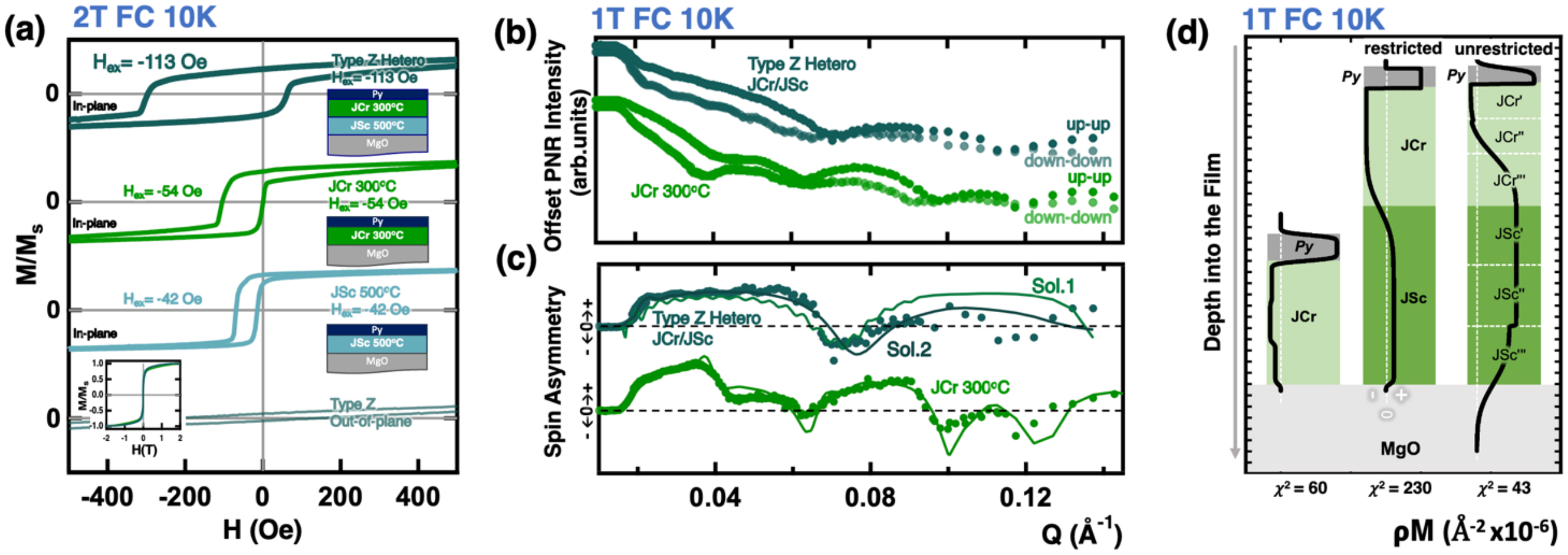

**Figure 6 | Exchange bias enhancement and depth-resolved magnetic structure in Type Z heterostructures.** (a) In-plane magnetic hysteresis loops measured at 10 K after field cooling in 2 T for Py-capped JSc (500 °C), JCr (300 °C), and the Type Z heterostructure (JCr/JSc), showing a pronounced enhancement of negative exchange bias in the heterostructure relative to the single layers. The lower panel shows out-of-plane measurements, confirming the absence of a ferromagnetic response, while the inset demonstrates magnetization saturation at high fields. Magnetization is normalized to the saturation magnetization shown in the inset. (b) Spin-dependent polarized neutron reflectivity measured at 10 K after field cooling in 1 T for JCr (300 °C) and the Type Z heterostructure, with vertically offset curves corresponding to different spin channels. (c) Spin asymmetry derived from the reflectivity data in (b), highlighting the distinct magnetic response of the heterostructure compared with the single-layer film. (d) Depth-resolved magnetic scattering-length-density (mSLD or $\rho_M$) profiles obtained from polarized neutron reflectometry fits, showing one representative solution for JCr and two solutions for the heterostructure. Both systems exhibit uncompensated interfacial spins at the Py/HEO interface, with a substantially larger buildup in the Type Z heterostructure. The lowest-$\chi^2$ solution for the heterostructure indicates a strong accumulation of uncompensated spins centered around the JCr/JSc interface. $\chi^2$ values indicate the corresponding fit quality.

To directly interrogate the depth-dependent magnetic structure underlying the enhanced exchange bias, we employ polarized neutron reflectometry (PNR) to quantify chemical and magnetic depth profiles of single layers and heterostructures. Figure 6(b) shows the spin-dependent neutron reflectivity for JCr and the Type Z heterostructure (both share similar Pt/Py/JCr at the top), measured in the polarized reflectivity spin-up and spin down channels. The corresponding spin asymmetry, defined as the normalized difference between these two reflectivity channels, is plotted in Figure 6(c) (details given in Methods) and reveals two distinct spin asymmetries. The analysis proceeds by inversely solving Schrödinger's equation for a trial scattering potential and iteratively refining the solution to minimize $\chi^2$ against the measured spin-dependent reflectivity, yielding a quantitative, depth-resolved magnetic scattering-length-density (mSLD) profiles Figure 6(d).

In the JCr single layer, mSLD becomes negative across the full film thickness, rather than being confined to the Py interface. This behavior indicates uncompensated spins-oriented opposite to the Py magnetization throughout the film, coupled with magnetometry, this suggests a globally antiferromagnetic state punctuated by localized unpercolated ferri- or ferro-like magnetic nanoregions (hence we do not observe them globally, Supporting Information Note 8 has M vs H for the case with No Py) that are consistent with the significant chemical and charge disorder in high entropy systems. The neutron data in Figure 6(b) and 6(c) suggests that Type Z heterostructure develops a qualitatively different magnetic configuration compared to JCr by itself. The first Type Z solution uses a restricted model where the JSc and JCr layers are treated as homogeneous layers with fixed chemistry and structural density. This model is physically most consistent with XRD and STEM analysis, it predicts significant residual unpaired spin at the permalloy interface and across the JCr layer (likely linked to strain-enhanced AFM exchange interaction), and it rationalizes the large exchange bias. While the fitting parameter is within all limits of acceptability, the uncertainties captured by $\chi^2 \approx 230$ could be reduced. To explore this possibility, a second unrestricted model is explored where each film is partitioned into three layers where the structural density in each is a free fitting variable that can include independent gradients. This solution yields a substantially improved fit quality ($\chi^2 \approx 43$); however, it introduces uncertainties regarding the heterostructure physicality, particularly concerning the effective structural density (Supporting Information Note 9). If true, the unrestricted solution predicts a pronounced buildup of uncompensated spins centered around the buried JCr/JSc interface and oriented in the same direction as the Py magnetization – a feature absent in the single-layer film. These results therefore provide two possible microscopic origins for the enhanced exchange bias in the Type Z architecture and a working hypothesis in which valence interfaces produce a distinct magnetic macrostate characterized by symmetry breaking and exchange-bias amplification that we believe is produced by and unique metastable macrostates in high entropy crystals.

These results raise a rich set of questions and opportunities that directly motivate future investigation, these include how individual layer thicknesses and the density of engineered interfaces govern the observed exchange behavior, and whether these parameters enable interfacial responses to dominate – and potentially supersede – bulk behavior in dense superlattices. Furthermore the high-entropy rock salt platform can accommodate many other highly misfit cations while maintaining anomalous crystallinity thus tunable framework for engineering coherent interfacial macrostates with strain, valence, magnetic contrast, and bandgap contrast[2,19,20,26–28]. More broadly, we view anomalous crystallinity as a new materials characteristic, associated with responsive chemical, defect, and structural disorders that compensate epitaxial and cation misfit strains to preserve exceptional crystallinity and texture and functionalities possibly beyond magnetism[10,13,14].

## Methods

### Bulk Targets and Thin Film Stack Synthesis

*Target synthesis*

Single-phase J14 and multiphase JSc and JCr ceramic ablation targets were prepared by reactive sintering of stoichiometric oxide powder mixtures: MgO (Sigma-Aldrich, 342793), CoO (Sigma-Aldrich, 343153), NiO (Sigma-Aldrich, 203882), CuO (Alfa Aesar, 44663), ZnO (Sigma-Aldrich, 96479), and $Sc_2O_3$ (Sigma-Aldrich, 294020) or $Cr_2O_3$ (Sigma-Aldrich, 393703). Powders were thoroughly mixed, milled with YSZ balls for 3 hours and sintered in air at 1000°C for 12 h, followed quenching in air at the cooling down direction when the furnace is at 750ºC.

*Thin-film growth.*

Thin films were deposited on MSE Supplies [001]-oriented MgO substrates by pulsed-laser deposition (PLD) using a Coherent 248 nm KrF excimer laser. Substrates were sequentially cleaned in acetone, isopropanol, and methanol, followed by a 25 min ultraviolet–ozone treatment, and then mounted to the heater buck using silver paint (Ted Pella Leitsilber 200). Substrate temperature was monitored by an optical pyrometer, as the heater thermocouple was found to systematically overestimate the true surface temperature. During growth, 50 sccm of $O_2$ was introduced into the chamber and the gate valve was throttled to stabilize the pressure at 50 mTorr. The laser fluence was 1.6 J $cm^{-2}$ (≈180 mJ per pulse), with a repetition rate of 10 Hz and 2,000 pulses per layer. Following deposition, all films were quenched to a load-lock within 2 min of the final laser pulse and immediately vented to ambient conditions, minimizing any post-growth annealing effects.

*Heterostructure growth*

For JSc/JSc heterostructures, an ~80 nm-thick rock-salt JSc layer was grown at 500 °C using 2,000 laser pulses. The substrate was then cooled to 300 °C, held for 30 min, and a second JSc layer was deposited using an additional 2,000 pulses. Growth conditions were chosen to maintain a constant deposition rate of ~24 nm $min^{-1}$, minimizing cation ordering and nanostructure formation. JSc/JCr heterostructures were grown following the same protocol, except that the target carousel was rotated during the cooldown to 300 °C to replace the JSc target with JCr for deposition of the top layer. Each layer was deposited using 2,000 laser pulses; under these conditions, the JCr layer reached a thickness of ~57 nm, corresponding to a growth rate of ~17 nm $min^{-1}$, less than that of JSc.

*Py deposition and VSM verification*

Approximately 7-10 nm of $Ni_{0.81}Fe_{0.19}$ (Permalloy, Py) was deposited onto the films by PLD at room temperature under 10 mTorr Ar. A 1 nm Pt capping layer was subsequently deposited in situ with same growth conditions used for Py to prevent oxidation. The laser was operated at a repetition rate of 6 Hz with a fluence of 1.2 J $cm^{-2}$. The magnetic response of the Py layer was verified using a Lake Shore vibrating sample magnetometer (VSM).

## X-ray Structural Analysis and Fast Reciprocal Space Maps

XRD was performed using a PANalytical Empyrean diffractometer. The primary incident beam optic was a hybrid mirror/2-bounce Ge monochromator combination; the primary diffracted beam optic was a programmable divergent slit and a PIXcel3D detector combination. The reciprocal space maps were collected in the "ultra-fast" mode where a set of two theta frames is collected for each continuous omega-2theta scan using frame-based 1D mode option available for a programmable divergent slit and a PIXcel3D detector combination.

## TEM Analysis

The TEM sample preparation was conducted using a Helios 660 SEM-FIB. Ion beams with 30 kV and 5 kV was used to extract the thin lamella, and a 2 kV beam was employed for the final thinning to achieve electron transparency. STEM experiments were performed on double Cs-corrected, monochromated Titan$^3$G2 microscope at MCL, Penn State, with an accelerating voltage of 300 kV. STEM images were acquired in orthogonal scan directions, and drift correction was carried out using open-source MATLAB code[29]. The monochromated EELS data were processed using HyperSpy package[30], and reference EELS spectra of Co were obtained from the literature[31]. Talos F200X was utilized to acquire the selected area electron diffraction and elemental distribution maps at an accelerating voltage of 200 kV. A Gaussian blur with a 1-pixel radius was applied to the EDS data during post-processing in Velox software, and the pixel intensities of each row were summed and plotted. Strain maps were generated via geometric phase analysis (GPA) of high-resolution STEM images using the Strain++ software package.[32] In this Fourier-space approach, local phase shifts from selected Bragg reflections were quantified relative to JSc thin film and converted to displacement fields and strain tensor components. The average Co oxidation state was determined from the L edge EELS spectrum via non-negative least-squares fitting to reference spectra, implemented in Python with scipy.optimize.nnls. The experimental spectrum was fit as a non-negative linear combination after energy-axis alignment, yielding fractional contributions that were summed to obtain the weighted average valence.

## SQUID Magnetic Measurements

Magnetization measurements were performed using a commercial Quantum Design MPMS3 superconducting quantum interference device (SQUID) magnetometer. Field-dependent hysteresis loops were collected with sweeping in-plane and out-of-plane magnetic fields up to 2 T using a field-cooling protocol to ensure a stable magnetic history between successive measurements at 10 K under field cooling. The diamagnetic background contribution from the substrate was determined by fitting the high-field region of the magnetization curves (above 1.5 T), where the film magnetization is saturated, to a linear function of applied field. This background signal was subsequently subtracted from the total measured moment to isolate the magnetic response of the thin film (More in supporting information Note 8).

## Polarized Neutron Reflectometery (PNR)

PNR was performed on the Magnetism Reflectometer at the Spallation Neutron Source at Oak Ridge National Laboratory. The instrument was configured for 30Hz operation utilizing a neutron wavelength spectrum of 2.2 Å to 8.2 Å. The neutron beam was polarized using a supper mirror polarizer. Access to the opposing spin-state is provided by a gradient-field RF flipper. The overall efficiency of the combined polarization system was above 96% of the wavelength range used[33]. The data was modeled using Refl1d[34] which can provide a Bayesian uncertainty envelope of the structural and magnetic scattering length density, SLD, mSLD or $\rho_S$, $\rho_M$, respectively. The calculation of the reflectivity first assumes a scattering potential for the neutron, $V(z) = 2ph^2/m_n [\rho_S(z) + \rho_M(z)]$, as a function of depth from the surface of the material. The software then solves Schrodinger's equation to calculate the reflectivity for each neutron spin-state as a function of momentum transfer, Qz. Additional fitting details are available in Supporting Information Note 9.

## Acknowledgements

The authors acknowledge the main support from the Penn State Materials Research Science and Engineering Center for Nanoscale Science under National Science Foundation award DMR-2011839. A portion of this research used resources at the Spallation Neutron Source, a DOE Office of Science User Facility operated by the Oak Ridge National Laboratory. The beam time was allocated to the Magnetism Reflectometer, BL4A, on proposal number IPTS-32857, IPTS-33292, and IPTS-35889. The authors would also like to acknowledge many helpful and insightful discussions with George N. Kotsonis, Long-Qing Chen, Venkatraman Gopalan, and Vincent H. Crespi regarding various aspects of this study.

## Supporting Information

**Note 1:** Anomalous Crystallinity

Anomalous Crystallinity has many manifestations that we highlight in the manuscript. To support and complement those manifestations we show below how, high solubility in thin films differ than bulk synthesis and we show comparison in thin film to low entropy counterparts.

(a) High solubility in thin films

Exemplary achievements of this approach include the incorporation of 16.7% $Sc^{3+}$ via far-from-equilibrium synthesis, which is at least four times higher than the bulk solubility limit. Figure S1 summarizes bulk ceramic synthesis of JSc performed at 1000 °C under near-equilibrium conditions, where $Sc_2O_3$ persists as a distinct secondary phase even at Sc concentrations as low as 5%. By contrast, far-from-equilibrium thin-film growth enables stabilization of significantly higher Sc contents within the rock salt lattice. Another example, not discussed in this manuscript, is the ability to incorporate highly misfit cations, such as 10% $Ca^{2+}$ despite a 38% ionic-radius mismatch, again without the formation of secondary phases[8].

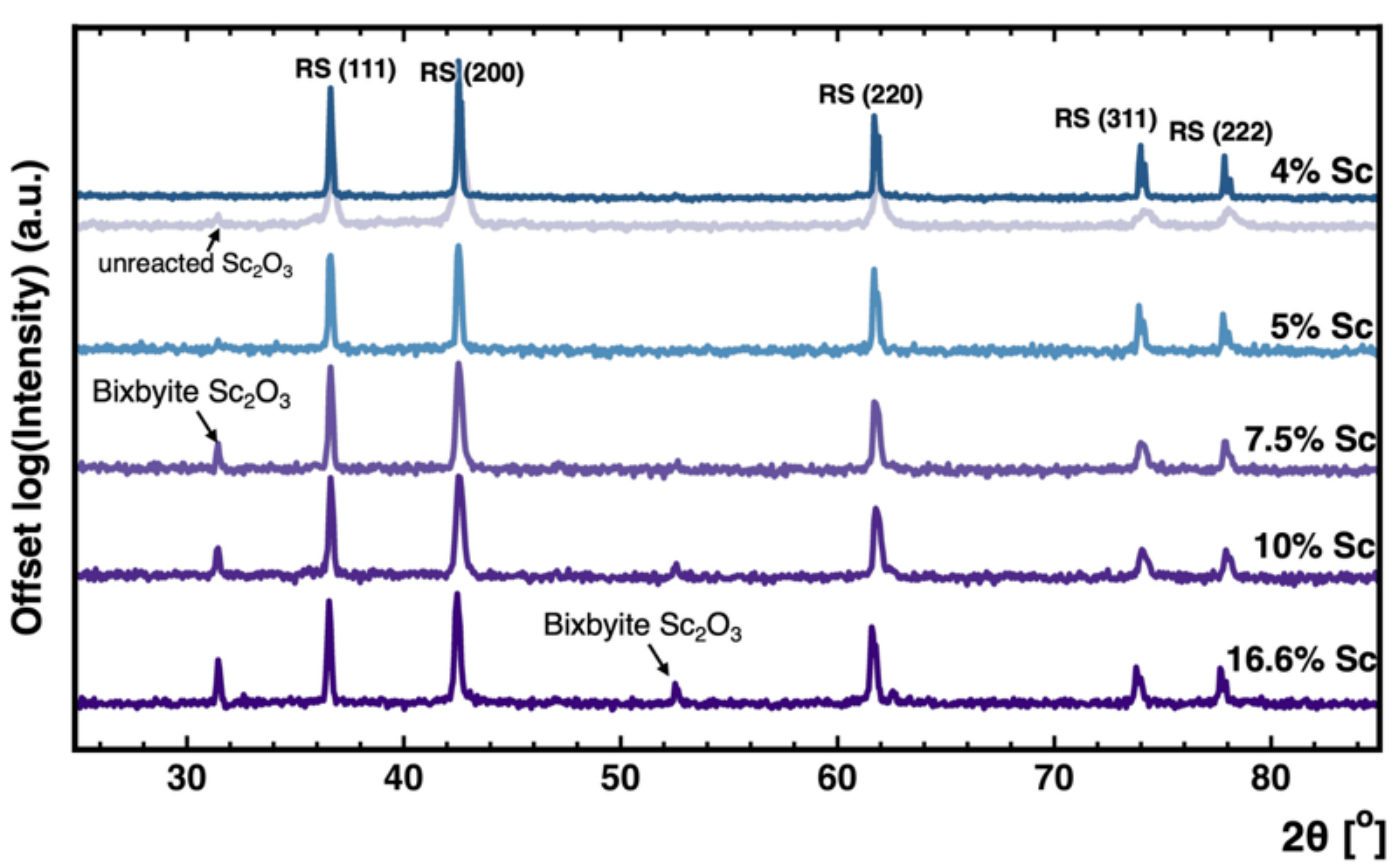

Figure S1: Figure S1. X-ray diffraction (XRD) patterns of bulk ceramic JSc synthesized at 1000 °C with varying Sc concentrations. Under near-equilibrium synthesis conditions, $Sc_2O_3$ (bixbyite phase) persists as a distinct secondary phase even at Sc contents as low as 5%, as indicated by the labeled reflections.

(b) Low entropy counterparts

It remains unclear whether the observed phase stability and tunability arise solely from far-from-equilibrium growth, or whether configurational entropy plays a decisive role. Complementary binary and ternary rock salt films grown under identical conditions serve as reference systems to isolate the role of configurational entropy in stabilizing the high-crystalline quality high-entropy lattice. The top panels of Figure S2 present BBHD XRD patterns centered on the (002) reflection, while the bottom panels show the same data collected using a crystal analyzer to more clearly resolve the film peaks. With the exception of MgCoCuO, which forms multiphase films, all compositions exhibit only the (002) reflection, confirming epitaxial growth on the substrate. Nevertheless, most binary and ternary films display substantially reduced crystalline quality compared to their six-component parent compositions, which consistently show sharp peaks and well-defined fringes. Importantly, even in well-grown cases such as MgNiO, the out-of-plane lattice parameter remains unchanged between 300 °C and 500 °C, indicating a loss of temperature-dependent lattice tunability. These results demonstrate that configurational entropy plays a critical role in the observed high crystalline fidelity and the tunable out-of-plane lattice responses.

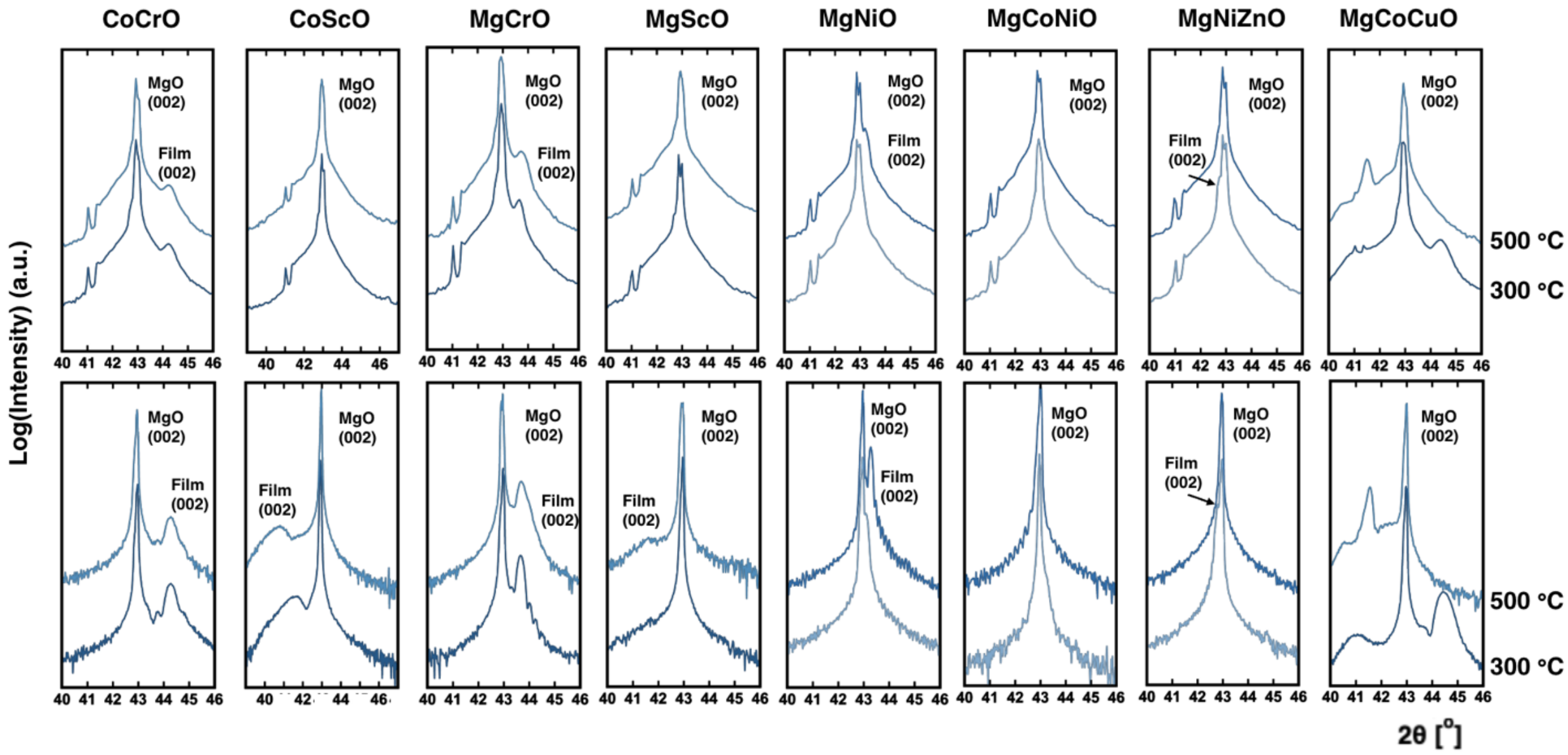

Figure S2. BBHD (Top) and High-resolution X-ray diffraction (bottom) 2θ-θ scans around the MgO (002) substrate reflection for a series of binary and ternary oxide thin films (CoCrO, CoScO, MgCrO, MgScO, MgNiO, MgCoNiO, MgNiZnO, and MgCoCuO). For each composition, data are shown for films grown at 300 °C and 500 °C, as indicated.

**Note 2:** X-ray diffraction and Reciprocal Space Maps of J14 with Co removed

In Figure S3, we examine the Co-free analogue MgNiCuZnO (denoted J–Co. In this case, the out-of-plane lattice parameter varies only weakly between 300 °C and 400 °C and remains essentially constant at higher temperatures, while the films remain fully coherent to the MgO substrate[2]. Unlike MgCoNiCuZnO, which switches between two distinct pseudo-strain states as a function of temperature, J–Co films remain pseudocompressively strained across the entire temperature range, with the in-plane lattice fixed to MgO and a comparatively expanded out-of-plane parameter. This supports our hypothesis that Co is one of the most important active players in controlling out-of-plane lattice parameter.

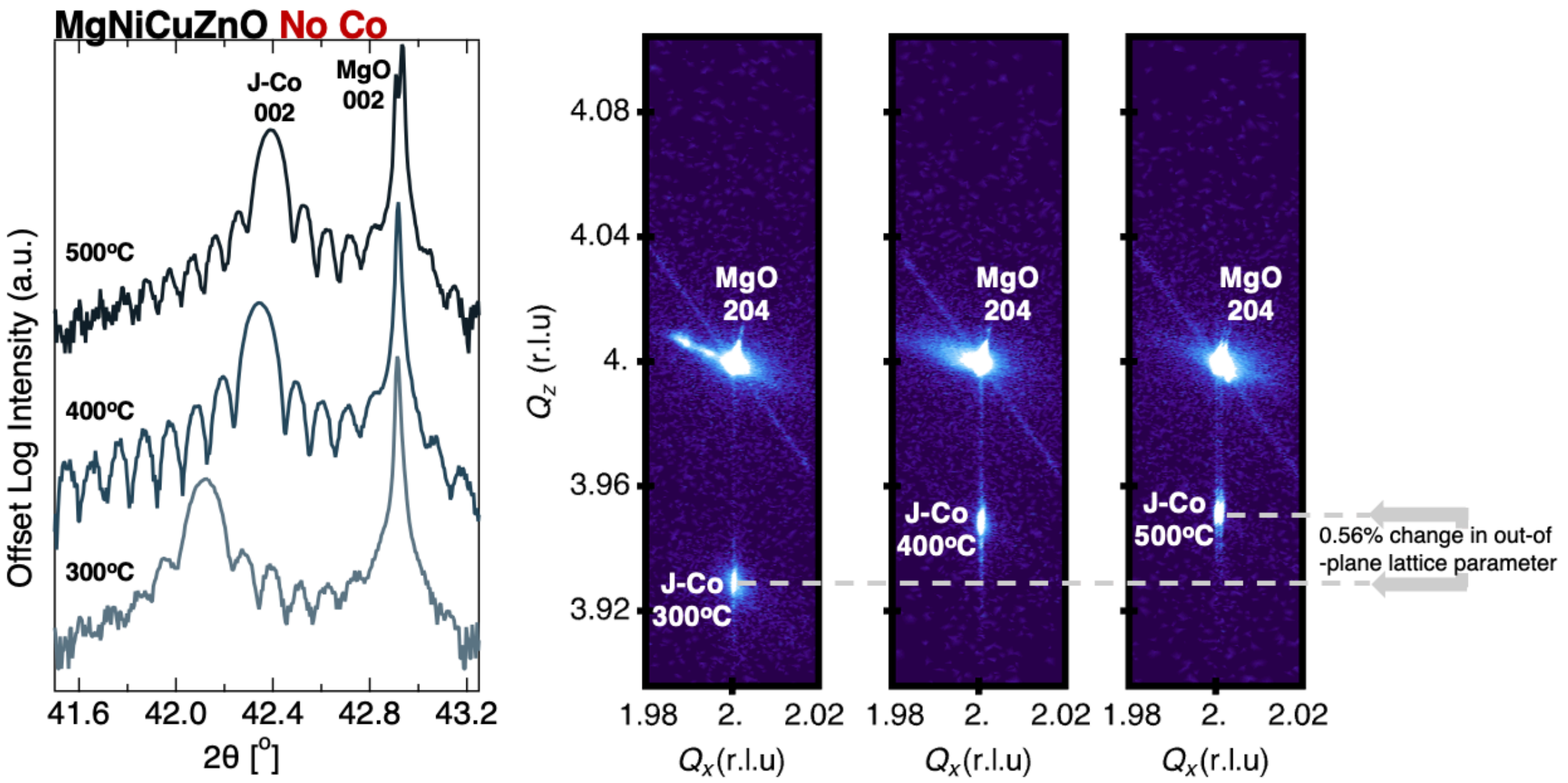

Figure S3: symmetric and asymmetric X-ray diffraction scans of MgNiCuZnO with no Co cations showing a small 0.56% change in out-of-plane lattice parameter with substrate temperature and all films are pseudomorphic to the substrate.

**Note 3:** Origin of double unit cell reflections in JCr grown at 500 °C using dark field TEM

Figure S4 summarizes the collected TEM data on JCr grown at 500oC. The double unit cell reflections in Figure S4(b) align well with spinel-like local symmetry/ordering.

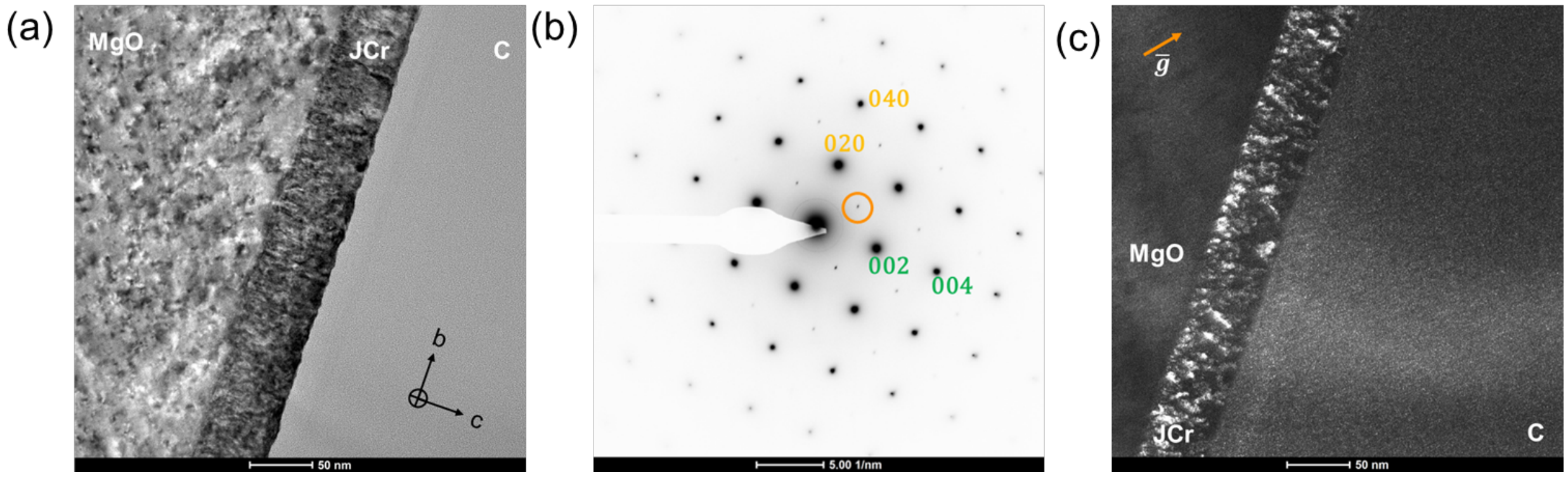

Figure S4: (a) Selected area region for electron diffraction shown in Figure 2b. (b) Corresponding electron diffraction pattern, with circular inset indicating the spinel reflection used for the dark-field TEM image in (c). (c) DF-TEM image revealing randomly distributed spinel regions throughout the thin film.

**Note 4:** Growth of JCr as a spinel on C-Saphire

To directly probe the epitaxial strain role in stabilizing the rock salt phase, we grew JCr films at 500°C on sapphire substrate, which impose an in-plane lattice parameter ~13% larger than MgO. XRD measurements Figure S5 reveal clear spinel reflections including the (111) peak near 18°, indicating bulk spinel phase formation. This contrast demonstrates that epitaxial strain, together with kinetic constraints during growth, suppresses full spinel conversion in JCr on MgO and strongly supports the lattice–valence framework developed here.

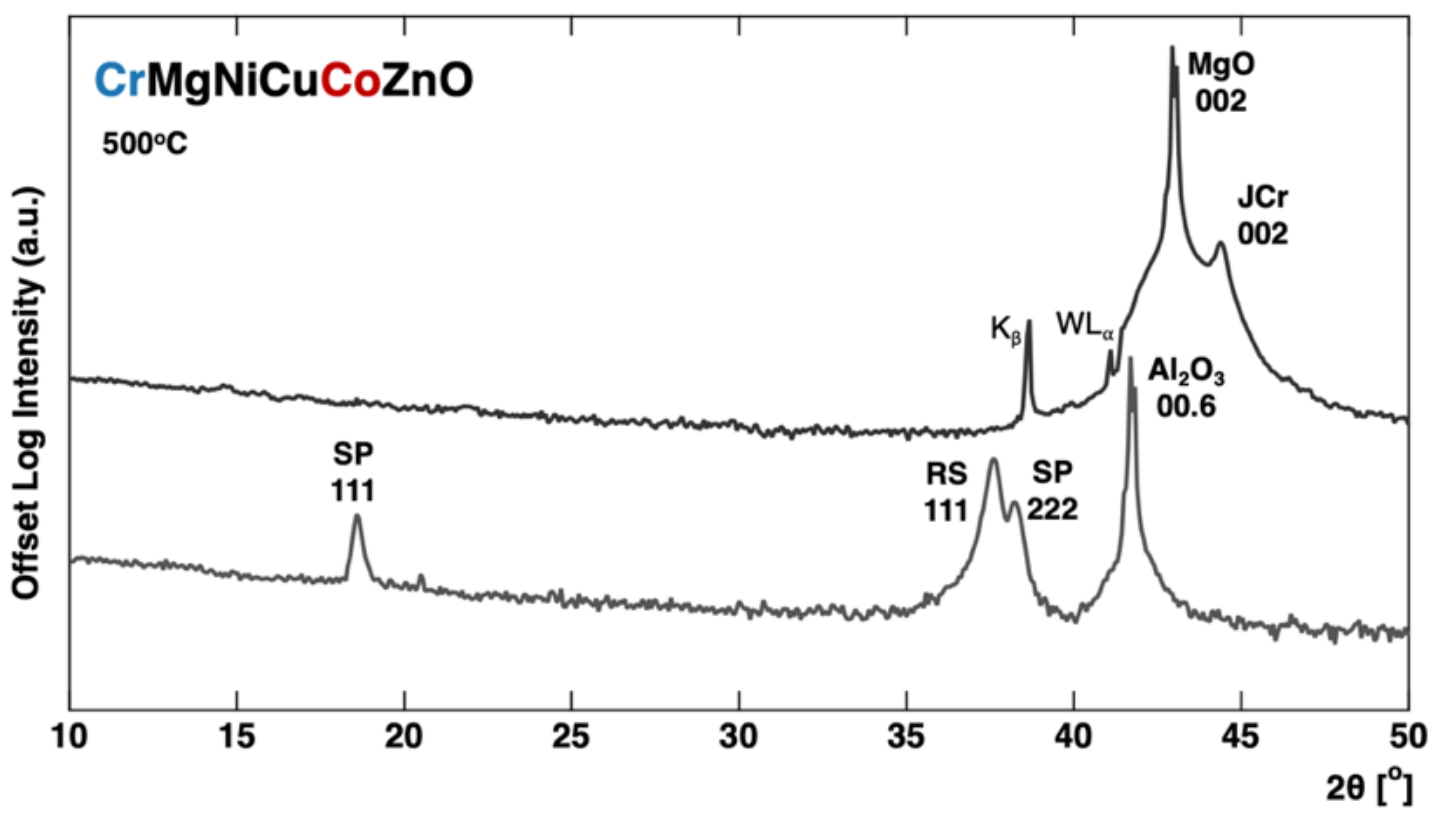

Figure S5: X-ray diffraction patterns of $Cr_{1/6}Mg_{1/6}Co_{1/6}Ni_{1/6}Cu_{1/6}Zn_{1/6}O$ (JCr) grown at 500oC on MgO and $Al_2O_3$ c-plane sapphire.

**Note 5:** Selected area region corresponding to electron diffraction in the heterostructures

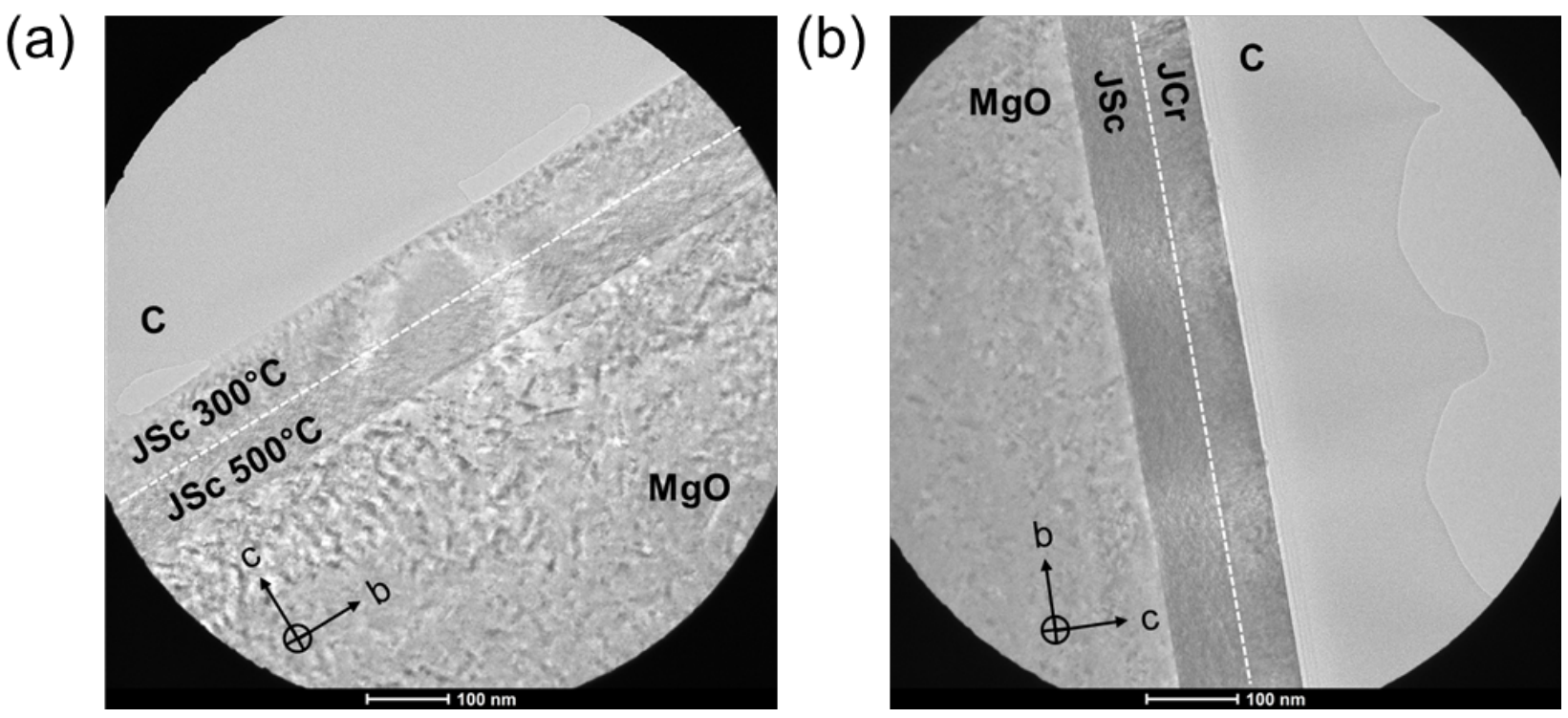

Figure S6: (a-b) Selected area regions for electron diffraction patterns shown in Figures 3c and 4c, respectively.

**Note 6:** An Example of Isochemical Type Z Heterostructure

We show in Figure S7 an example where layers are still isochemical and heterovalent, but the heterovalency is sufficient to switch the pseudostrain state from tension to compression relative to the substrate. This structure is achieved by growing two J14 layers while decreasing the substrate temperature during crystal growth, analogous to the method used for growing Type X. The resulting heterostructure consists of two epitaxial rock salt layers with distinct lattice parameters and a well-defined interface separating them. One layer has an out-of-plane lattice parameter larger than that of MgO, while the other has a smaller out-of-plane lattice parameter than that of MgO. Pendellösung fringes indicative of smooth interfaces are observed and the reciprocal space map indicates that both layers are commensurate with the MgO substrate, having in-plane lattice parameters of 4.21 Å with 3.15% change in out-of-plane lattice parameter.

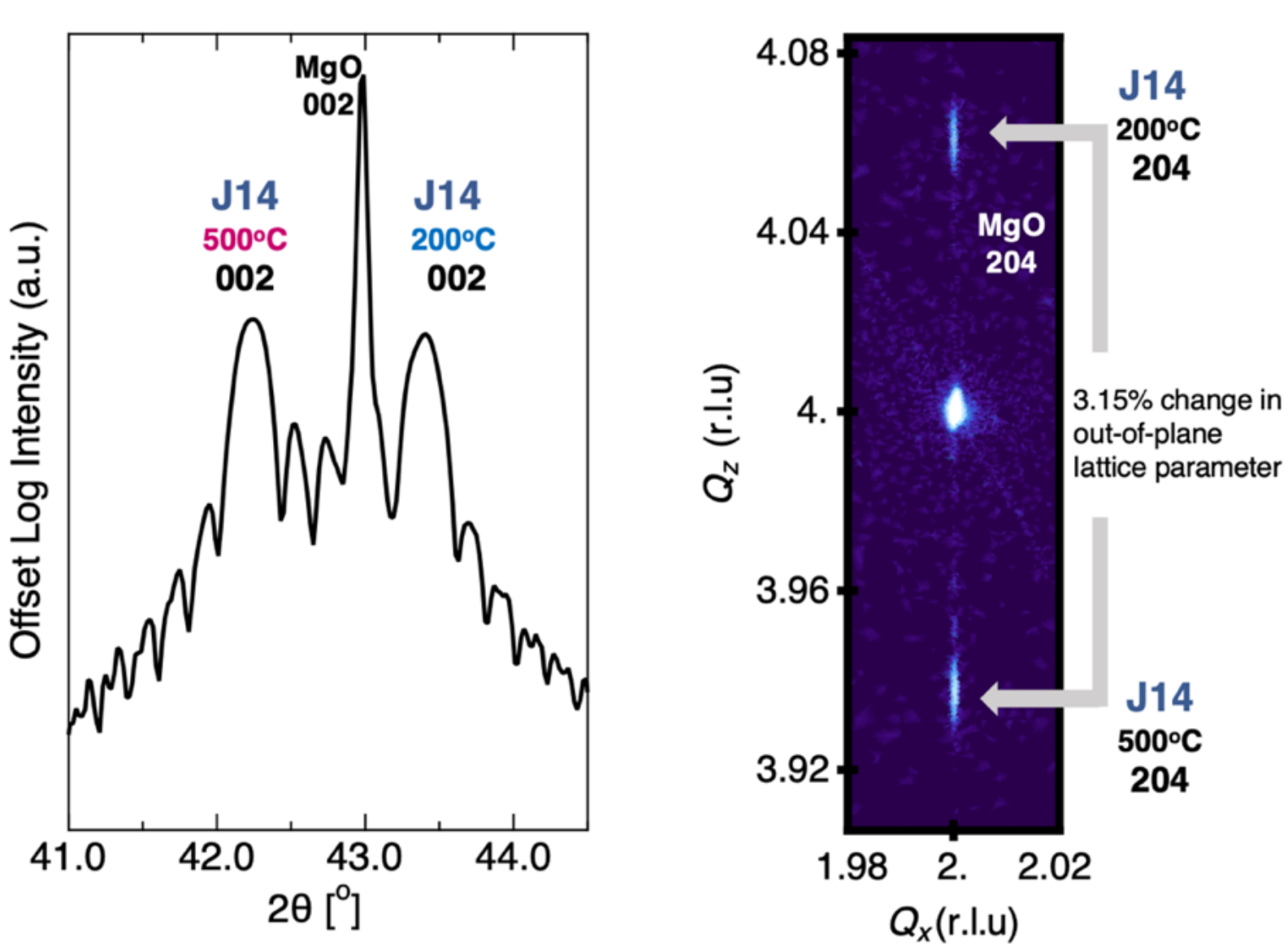

Figure S7: Structural characterization Isochemical Heterostructure Tyze Z with 3.15% change in out-of-plane lattice parameter: (a) XRD pattern; and (b) the reciprocal space map of the heterostructure.

**Note 7:** Strain maps of JCr/JSc stacked samples using geometric phase analysis (GPA)

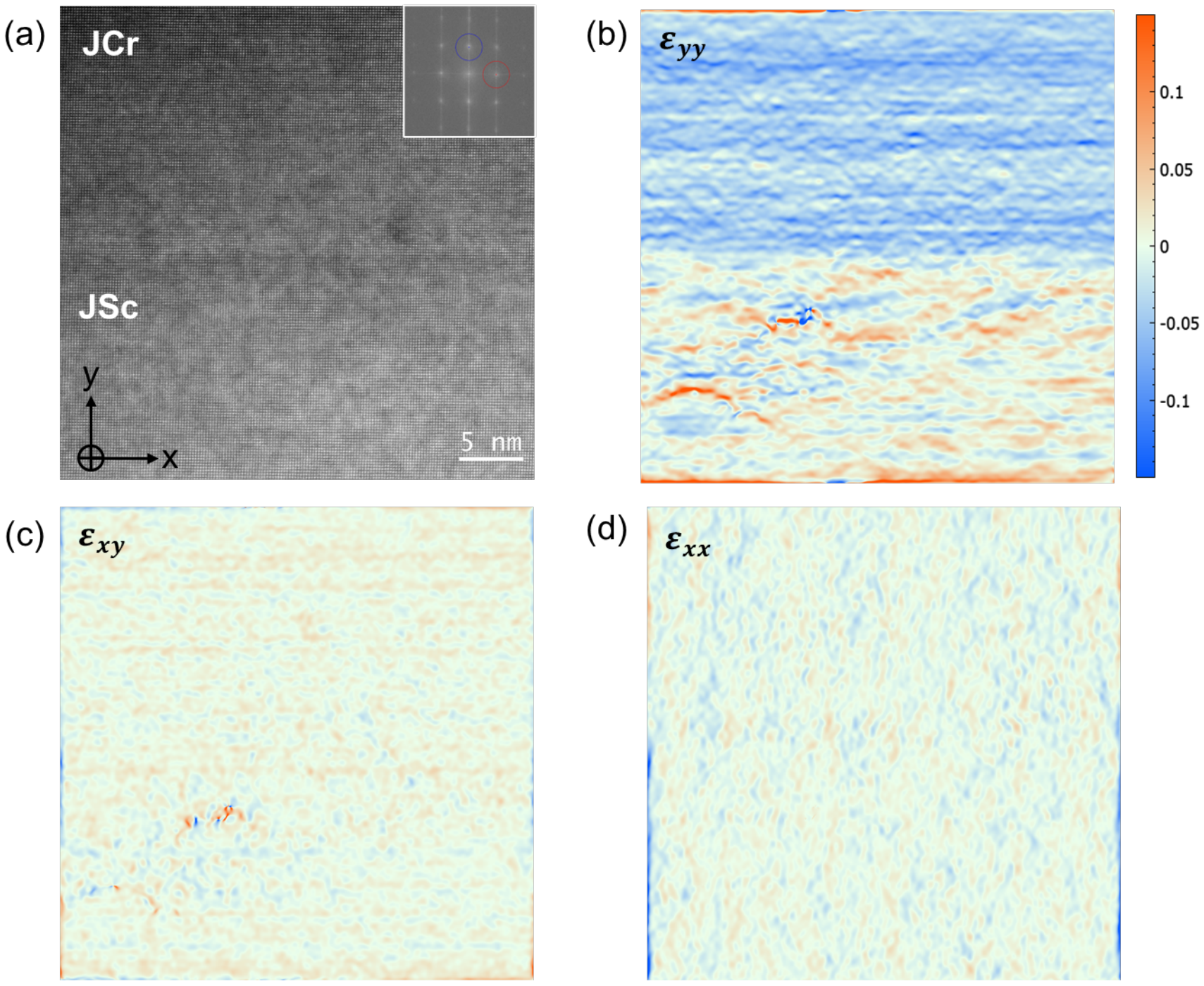

Figure S8: (a) STEM image with inset showing the FFT and Bragg vectors used for GPA analysis. (b-d) Corresponding strain maps $\varepsilon_{yy}$, $\varepsilon_{xy}$, $\varepsilon_{xx}$, respectively.

Strain maps were obtained by geometric phase analysis of the cross-sectional STEM images using the Strain++ software, taking a region of the JSc thin film as the strain-free reference region. The out-of-plane strain component $\varepsilon_{yy}$ clearly reveals that the JCr thin film experiences compressive strain in the out of plane relative to the JSc reference, which confirms the pseudotensile strain state. In other words, the JCr film is smaller than MgO so it gets stretched in plane in pseudotensile state which results in it being compressed out of plane relative to JSc. In contrast, the in-plane strain component $\varepsilon_{xx}$ shows no discernible difference between JCr and JSc films. This demonstrates that the in-plane lattice parameter remains effectively pinned to the underlying substrate across both compositions.

**Note 8:** SQUID control experiments: measurements for MgO substrate, and Type Z heterostructure with and without Py

Figure S9 summarizes the SQUID magnetometry measurements performed after field cooling in 2 T to 10 K for three cases: the bare MgO substrate, the Type Z heterostructure without Py, and the Type Z heterostructure with Py. This control experiment serves as a null test to strengthen confidence in the magnetic signal observed for the Type Z with Py on top sample. Importantly, within the resolution of our measurements, the Type Z heterostructure without the Py overlayer exhibits no detectable ferromagnetic response, confirming that the observed magnetization in the full stack arises from the presence of Py and its field cooling within the complete FM/AFM/AFM/substrate architecture.

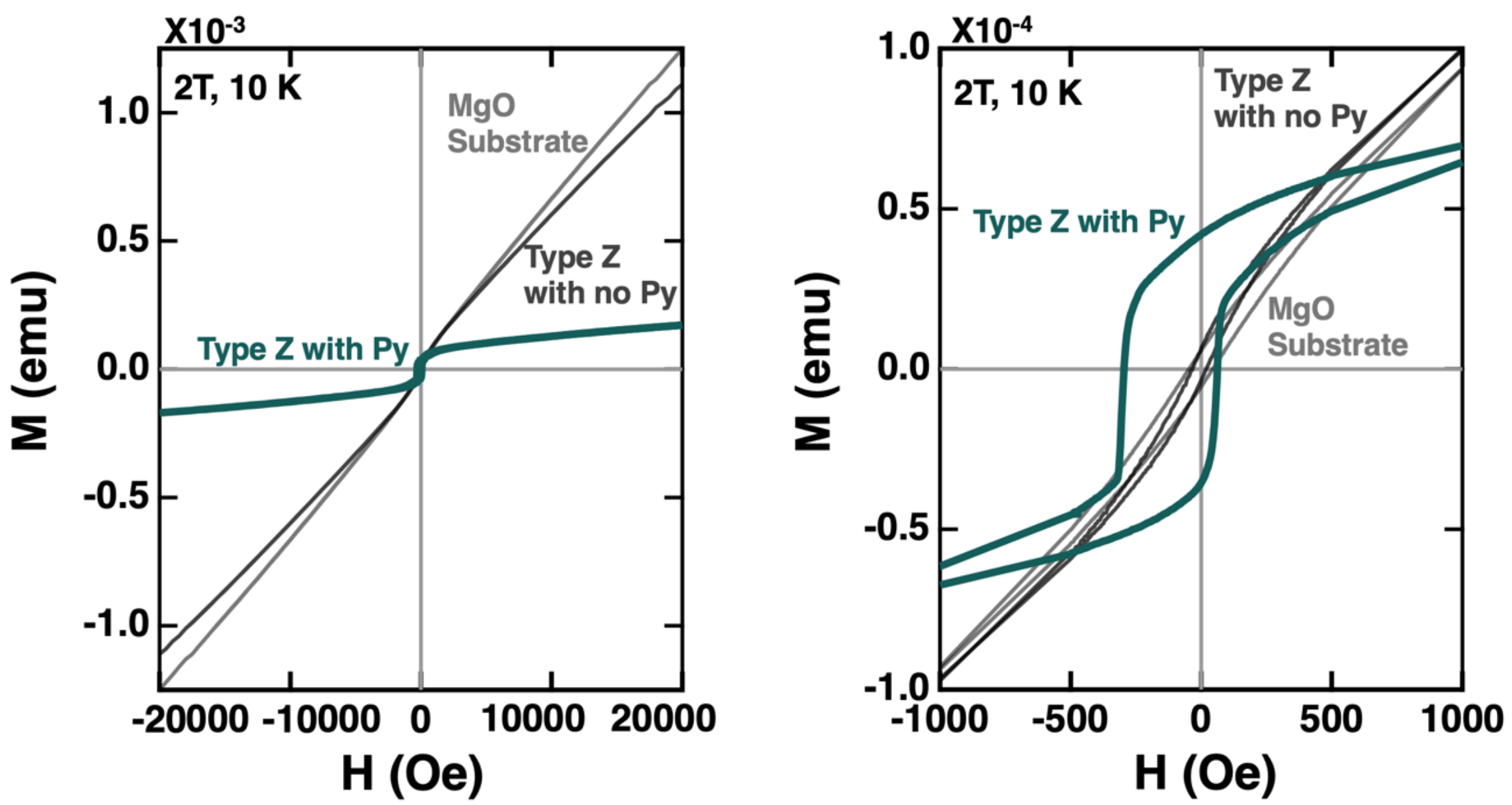

Figure S9: Magnetization–field (M–H) loops measured at 10 K after field cooling in 2 T for three configurations: a bare MgO substrate, the Type Z heterostructure without a Py overlayer, and the complete Type Z heterostructure with Py.

**Note 9:** Other Magnetic Fields and Error Quantification in PNR Measurements

For Type Z heterostructure PNR, we employed a global fitting strategy in which a single structural model was constrained to simultaneously reproduce the data acquired at all four applied magnetic fields (5 Oe, 175 Oe, 350 Oe, and 1 T), as highlighted in Figure S10. The fit shown in the main text corresponds to the 1 T dataset; however, the conclusions are drawn from fits that consistently satisfy the full field-dependent dataset.

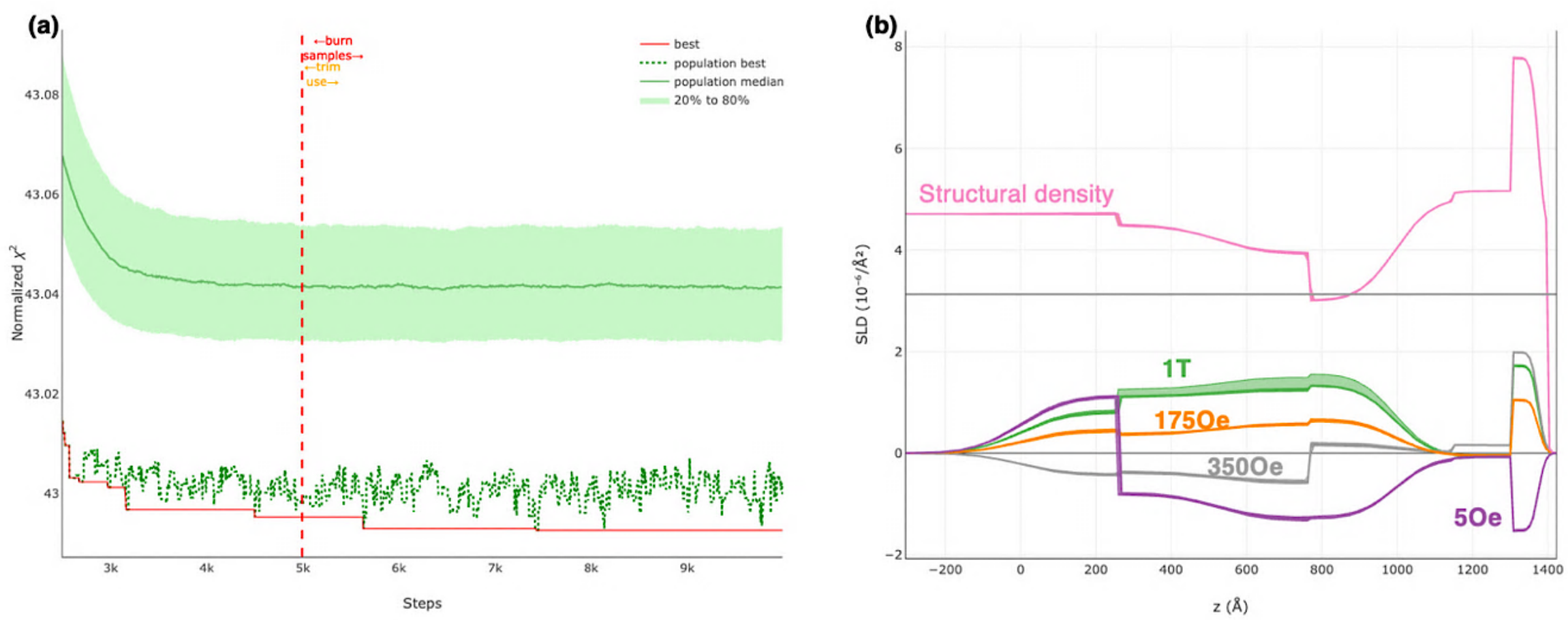

Figure S10. (a) Convergence plot for Type Z heterostructure fit generated by REFL1D. (b) Corresponding structural and magnetic scattering length density (SLD) profiles, including uncertainty dispersion, corresponding to the burn-in sample ensemble highlighted in (a).

Across all applied fields, the fits reveal the emergence of uncompensated interfacial spins at the JSc/JCr interface, indicating a robust interfacial magnetic reconstruction. While this magnetic profile could be physically consistent with accommodation mechanisms arising from the ~5.5% out-of-plane lattice mismatch, including strain gradients that may enhance spin frustration and stabilize uncompensated spins around the JCr/JSc interface, the physical manifestations of structural density variation are less clear. Additionally, the sign of the interfacial magnetization depends on the applied field: the 5 Oe and 350 Oe datasets exhibit a negative interfacial spin polarization, whereas the 175 Oe and 1 T datasets show a positive polarization. While the origin of the negative interfacial moment at 350 Oe remains unclear and warrants further experimental investigation, we emphasize that the qualitative presence of uncompensated interfacial spins is a reproducible and field-independent feature of the fits. An additional point of interest is that, at 5 Oe, the Py layer itself exhibits a negative magnetization relative to the higher-field measurements. This behavior may correlate with the asymmetric coercive fields observed in the JSc/JCr heterostructure, where opposite signs of coercivity are present in the magnetization vs field data.

All parameters, constraints, and model definitions required to reproduce the PNR fits will be made available through a shared data repository provided by the publishing journal or directly by the corresponding author.